\documentclass{article}

\newcommand{\ket}[1]{{|#1\rangle}}

\author{Christof Zalka \\
Department of Physics, University of Waterloo \\
Waterloo, Ontario, Canada N2L 3G1}

\title{Implementing high dimensional unitary representations of SU(2) on
a Quantum Computer}

\begin{document}
\maketitle

\begin{abstract}
In this note we consider a system with a large angular momentum $\ell$
whose state we can store using some $\log_2(\ell)$ qubits. The problem
then is how to carry out spatial rotations of the system in this
representation. In other words we are looking at a unitary
representation of SU(2) with dimension $2 \ell+1$ and want to
implement these transformations with resources polynomial in
$\log(\ell)$. We only give a sketch of our solution which involves
``storing'' discretised spherical harmonic functions $Y_{\ell
m}(\Theta,\phi)$ in a quantum register.  Also there are some technical
gaps in the construction, but they are based on plausible assumptions.
Our approach is rather cumbersome and we hope somebody will find a
nicer solution. For a nice, elementary explanation of what we are
trying to do (not involving physics or representation theory) see
section \ref{symm}.
\end{abstract}

\section{Introduction}
Consider a system (say a particle) with total angular momentum given
by $\ell$, thus the absolute value squared of the angular momentum is
$\ell(\ell+1)$. The $2 \ell+1$ dimensional state space (Hilbert space)
of this system is spanned by the eigenstates of the z-component of the
angular momentum with eigenvalues $m=-\ell\dots+\ell$. (For the time
being we only consider integer $\ell$, thus e.g. orbital angular
momentum in quantum mechanics.)  We can now easily represent a (pure)
state of this system in a quantum register with some $\log_2(\ell)$
qubits, whereby we choose the m-eigenstates to correspond to the
computational basis states: $ \ket{m},~ m=-\ell \dots +\ell $.

A rotation about the $z$-axis is easily implemented as $\ket{m} \to
e^{i m \phi} \ket{m}$. If we knew how to implement some (actually
almost any) other spatial (thus SO(3)) rotation we could combine it
with $z$-rotations to obtain any rotation. But doing such a rotation
has proved quite tricky and is the main point of this note.  In more
mathematical terms we are simply trying to efficiently implement the
unitary transformations corresponding to a high dimensional
representation of SU(2) (or SO(3)) on a quantum computer.

The approach we present here is rather cumbersome but it seems to do
the job of efficiently implementing any rotation with arbitrarily
small (exponentially small) error. Still there are some gaps, but we
believe these are just (tedious) technicalities. Also, as mentioned
before, we hope that someone finds a much nicer (possibly exact)
solution, although this has eluded us.

\section{The approach}
The idea is that smooth functions in 3 dimensions (thus functions from
$R^3 \to C$) form an SO(3) representation under spatial rotations. In
particular the spherical harmonics $Y_{\ell m}(\Theta,\phi)$ for a
fixed $\ell$ transform under spatial rotations like the angular
momentum eigenstates $\ket{\ell,m}$, thus they form an irreducible
representation of SO(3) (and thus also of SU(2)). (Thus a rotated
spherical harmonic can be written as a linear combination of $Y_{\ell
m}$'s with the same $\ell$.)  Thus in a way we are simply looking at a
wavefunction of a particle in 3 dimensions which has the right
(orbital) angular momentum. (Because for the angular momentum the
radial dependence doesn't matter, we can actually look at
wavefunctions which depend only on $\Theta$ and $\phi$.) For a nice
picture of a random superposition (drawn from a Gaussian distribution)
of spherical harmonics with $\ell=20$ look at
\verb|http://www.zalka.itp.unibe.ch/ylm.jpg| or
\verb|http://www.iqc.ca/~zalka/LANL/graph/ylm.jpg| ~.

The idea now is to represent (``store'') such functions on the quantum
computer in a discretised form using a sufficiently fine 3-dimensional
cubic lattice. Thus we have 3 quantum registers $\ket{x,y,z}$ which we
prepare in a superposition with amplitudes equal to the function value
at the point $x,y,z$ (apart from an overall normalisation). Still, we
may ultimately like to apply the (representation of the) rotations
to the compact encoding in a single register $\ket{m}$. For this we
first have to translate this encoding into the more complicated one
representing a 3-dimensional function. Thus to implement a rotation we
have to do the following 3 (unitary) steps:

\begin{itemize}
\item translate: $\ket{m} \to \ket{Y_{\ell m}}$
\item rotate:  spatially rotate $\ket{Y_{\ell m}}$
\item translate back: $\ket{Y_{\ell m}} \to \ket{m}$
\end{itemize}

Where everything should just as well work for superpositions over
different $m$'s. The state $\ket{Y_{\ell m}}$ stands for a suitably
discretised version of the function $Y_{\ell m}(\Theta,\phi)$ on a
3-dimension grid in the 3 registers $\ket{x,y,z}$. The discretisation
has to be fine enough to faithfully represent the smooth function, but
this can be done with only $O(\log(\ell))$ qubits. We also have to
choose some $r$-dependence and I propose to simply consider the
spherical harmonics on a thin spherical ``shell''. Thus really we have
the overall function $Y_{\ell m}(\Theta,\phi) \cdot R(r)$ where $R(r)$
is only non-zero in a narrow region around some $r_0$ where e.g. we
choose it to be constant.

One way to do the ``translation'' is to first think how we could
prepare the state $\ket{Y_{\ell m}}$ for a fixed $m$. If we knew that
(and I will sketch how that could be done), we can implement the
unitary transformation $\ket{m} \to \ket{m,Y_{\ell m}}$. To see how we
then can get rid of the $m$, consider the converse: given
$\ket{Y_{\ell m}}$, how can we find $m$? For this, note that a spatial
rotation by $\phi$ about the $z$-axis of $Y_{\ell m}$ gives an overall
phase of $e^{i m \phi}$. Thus, if we know how to spatially rotate
$\ket{Y_{\ell m}}$, we can do ``phase estimation'' (see Kitaev, Mosca)
to do $\ket{Y_{\ell m}} \to \ket{Y_{\ell m},m}$. Combining these
techniques approximately gives a unitary transformation doing $\ket{m}
\to \ket{Y_{\ell m}}$.


How can we do the spatial rotation of $\ket{Y_{\ell m}}$ ? Of course
rotations by 90$^o$ about the principal axes are easy as the points on
the rotated grid (a cubic lattice) match those of the original one,
e.g. we can do $\ket{x,y,z} \to \ket{x,z,-y}$ reversibly and thus
unitarily. For rotations about some generic angle we have to find a
1-to-1 mapping between the points of the rotated and the original
lattice. Small errors, thus mapping a point to another point a few
lattice spacings away won't matter, as we can make the lattice spacing
exponentially small and thus the effective error will be
small. Unfortunately simply mapping a point to the nearest point of
the other lattice isn't 1-to-1 and thus not unitary. Thus we have to
think of something different (see below).

\section{(some) techniques}
%

\subsection{rotation on a lattice}
What we need is a 1-to-1 mapping between points of two cubic lattices
which are rotated with respect to each other. As it is sufficient to
do rotations about one of the principal axes, the problem really
reduces to two 2-dimensional (square) lattices with the same
origin. The idea is to do several ``shearings'' which together give a
rotation. A ``shearing'' is a linear map of the form $x \to x+ a\cdot
y,~ y \to y$, thus a mapping where all lines parallel to one of the
axes are mapped into themselves. Doing 3 such mappings:
$$ \left( \begin{array}{cc} 1 & a \\ & 1 \end{array} \right)
\left( \begin{array}{cc} 1 & \\ b & 1 \end{array} \right)
\left( \begin{array}{cc} 1 & a \\ & 1 \end{array} \right) $$

we can e.g. determine $b$ in terms of $a$ such that overall we get a
rotation (this is an easy exercise). The individual ``shearings'' can
rather easily be done, as they simply map lines into themselves with a
translation. We can then easily compute a map (and its inverse) that
e.g. maps each point on a line to the next one of the other lattice.

But the lines are finite (as the lattice is finite). Then, say, we use
periodic boundary conditions, thus our mapping is cyclic. At any rate,
the outer parts of the whole lattice will not be mapped according to a
rotation, this only happens (approximately) for a region around the
centre of the lattice. But this is no problem, as we simply make sure
that our wavefunction is only non-zero in this inner region.

\subsection{preparing spherical harmonic states $\ket{Y_{\ell m}}$}
Thus we want to prepare a state (of a quantum register) whose
amplitudes are taken from a smooth, ``well known'', natural
function. Let's first consider the 1-dimensional case. Thus we have a
``natural'' function $f(x)$ and want to prepare the state $\sum_x f(x)
\ket{x}$ (assuming this is normalised) where, say, $x=0 \dots
2^n-1$. In this state the probability of finding the highest order bit
in state $\ket{0}$ is $\sum_{x=0}^{2^{n-1}-1} |f(x)|^2$, which we can
achieve by suitably rotating (say with an SO(2) rotation) a qubit
initially in state $\ket{0}$. To get the right probabilities for the
TWO high order bits we have to suitably rotate the second highest
qubit (initially in state $\ket{0}$). First we have to determine the
sums over the 4 subintervals $\sum_{x=a}^{a+2^{n-2}-1} |f(x)|^2$ with
$a=0,2^{n-2},2 \cdot 2^{n-2}$ and $3 \cdot 2^{n-2}$. Depending on the
values of these 4 sums and conditioned on whether the highest bit is 0
or 1, we have to rotate the second highest qubit. And so on. Thus at
one point we have to rotate qubit number $i$ by an angle which depends
on the values of all higher bits (and on the function). The correct
rotation angle will be given by the ratio of two sums of the form
$\sum_{x=k\cdot 2^i}^{(k+1) \cdot 2^i-1} |f(x)|^2$ with $k=k_0$ and
$k=k_0+1$.

Thus before carrying out the controlled SO(2) rotation of qubit number
$i$, we have to compute the value of two of these sums (which for
smooth functions can be approximated by integrals). This computation
has to be carried out ``in quantum parallelism''. Thus the value of
the higher order bits determines which sums we have to compute. This
(approximative) computation of integrals is the most cumbersome part
of our construction. Still, for ``natural'', ``well known'' functions
like the spherical harmonics, we can expect to find an efficient
algorithm for this. In the end (if we use conditional SO(2)
rotations), we get a state that has the right amplitudes, possibly up
to phase. But a ``rephasing'', again according to a well known,
natural function is rather easy to carry out efficiently. This whole
technique is e.g. described in \cite{sim}, section 3.2 (probably more
carefully than here...) and has also been described by other authors.


Clearly the technique also generalises to functions of several
variables (leading to a state of several quantum registers). Instead
of integrals over ``binary intervals'' we will now have to compute
integrals over ``binary hyper-intervals''. In our case we have the
function $Y_{\ell m}(\Theta,\phi) \cdot R(r)$ on a 3-dimensional
lattice. In a first step we could get the right probability
distribution over z-values. Thus in the registers $\ket{x,y,z}$ we
would have $x=y=0$ and thus a distribution concentrated on the
z-axis. For this first step we would only have to consider the
generalised Legendre functions $P_{\ell m}(z=\cos(\Theta))$ and
integrals over it. (Remember that $Y_{\ell m}(\Theta,\phi) = P_{\ell
m}(\cos(\Theta)) \cdot e^{i m \phi}$.) From there all we would have to
do is (thinking of one $z=const.$ plane) to ``spread out'' the $x=y=0$
point into a ring and do a rephasing according to $e^{i m
\phi}$. (It's a ring because we chose a narrow, rectangular radial
function $R(r)$.)

The main technical gap in our construction is to show how integrals
over the generalised Legendre functions $P_{\ell m}(z)$ can be
computed (approximatively) efficiently. I believe it is clear that
this is possible!

\section{various remarks}

\subsection{How to measure ${\vec L}^2$ ?!}
We have shown how $m$ (z-component of the orbital angular momentum)
could be measured for a wavefunction on a lattice. But I don't know
how to measure $\ell$, thus the absolute value squared of the angular
momentum. ${\vec L}^2 = L_x^2 + L_y^2 + L_z^2 =
\Delta_{\Theta,\phi}$. But even though we know how to measure $L_x^2$,
$L_y^2$, $L_z^2$ (and also $\Delta_{x,y,z}$) this doesn't allow us to
measure ${\vec L}^2$, because these terms don't commute. If we knew
how to measure ${\vec L}^2$, this would allow us to initialise the
$\ket{x,y,z}$ register in a specific irreducible representation of
SO(3), although maybe not one of our choosing. Still, for many
applications this would be enough and would relieve us from doing the
cumbersome $\ket{m} \to \ket{Y_{\ell m}}$ transformation. Failing
other ways, one way of measuring ${\vec L}^2$ might be to do a ``full
translation'' $\ket{\ell,m} \to \ket{Y_{\ell m}}$ (involving
estimating $\ell$ from $\ket{Y_{\ell,m}}$) but then that would be
again very cumbersome.

\subsection{Doing without the ``compact representation'' ?}
As mentioned above, sometimes we may not need to apply the
transformation to the ``compact'' $\ket{m}$ representation, just
applying it to the 3D lattice representation might be good
enough. (E.g. if all we need is some kind of ``phase kickback'', it
doesn't matter what representation we have.) Still, we may want to
initialise things if not in a specific state, then at least in a
specific irreducible representation. But, as mentioned above, we
haven't found an easy way of doing this.

\subsection{Doing things on a 2D $\Theta, \phi$ lattice.}
There are alternatives to representing the spherical harmonics
$Y_{\ell m}(\Theta,\phi)$ on a 3-dimensional cubic lattice. E.g. we
could simply use a 2-dimensional $cos(\Theta),\phi$ plane (a square
lattice). The problem is that the spatial rotations then get more
complicated. E.g. a rotation by $45^o$ about the $x$-axis (which would
be sufficient) would correspond to some smooth (but not linear)
mapping of this plane into itself. Again we could try to work out how
to make a 1-to-1 mapping to close points (the ``smooth mapping'' is
area preserving, thus the density of points remains unchanged),
possibly by piecewise linearising the smooth map.

An advantage of using the $cos(\Theta),\phi$ plane would be that we
wouldn't have the problem of uncomputing $m$: $\ket{Y_{\ell m},m} \to
\ket{Y_{\ell m}}$. We could first construct $P_{\ell, m}(z)$: $\ket{m}
\to \ket{P_{\ell, m}(z),m}$ and then simply Fourier transform $m$ to
get the correct $\phi$-dependence. Possibly similar techniques
(avoiding the ``phase estimation'') are also possible for the 3D
lattice.

\subsection{``Changing lattices'': useful for quantum simulations.}
So we represent a (smooth) wavefunction on a lattice in some
coordinate system. Sometimes we may want to change the lattice,
e.g. by rotating it, as we did here, or e.g. by changing from a
Cartesian lattice to a spherical coordinate one (with same density of
lattice points). This is similar to a problem in image processing when
we want to change the grid of pixels (we may call this
``rescanning''). But here, in order to do things unitarily, we have to
find a 1-to-1 mapping between close points. (Unless we want to do
thing non-classically, thus mapping a point on one lattice to a
superposition of nearby points of the other lattice.) E.g. if we want
to measure the $z$-component of the angular momentum, a polar
coordinate lattice would be nicer. (This would e.g. give a nicer way
of estimating $m$ from $\ket{Y_{\ell,m}}$.)

\subsection{Why don't ``efficient quantum simulations'' solve our
problem ?}
E.g. Seth Lloyd (Science, August 1996) (but also \cite{sim}) has shown
that physical quantum systems can efficiently be simulated on a
quantum computer. One may wonder whether that would not already
include what we are doing here, namely ``efficiently simulating large
angular momenta''. The point is that it has been shown how to do
simulations that are polynomial (actually linear) in the physical size
of a quantum system, thus space, time, energy, precision, etc. But
although we can store the state of an angular momentum $\ell$ using
only $O(\log(\ell))$ qubits, the physical size (here space, energy) of
such a system is necessarily large, namely polynomial in $\ell$. Thus
what we are trying to do is somewhat more ambitious than an
``efficient quantum simulation''.

Apart from angular momentum one may think of simulating other
``few-particle'' system with high dimension, e.g. two interacting
harmonic oscillators (truncated at some high energy).

\subsection{``Adding'' angular momenta}
``Angular momentum addition'' is a standard subject in quantum
mechanics (involving Glebsch-Gordon coefficients etc.). Two systems
(thus formally a tensor product) with angular momentum quantum numbers
$\ell$ and $\ell'$ carry irreducible representations $\ell-\ell'$ to
$\ell+\ell'$ (assuming $\ell \geq \ell'$). Thus in a way if we have
representations $\ell$ and $\ell'$ we have in particular also
$\ell+\ell'$. It is also not difficult to find a ``translation'' from
a compact encoding of the large representation (on a QC) to the two
smaller ones: $\ket{M} \to c \cdot \sum_m \ket{m,m'=M-m}$. Thus if we
know how to simulate $\ell$ and $\ell'$, then we can also do
$\ell+\ell'$.

This is why our problem would already be solved even if we could only
``produce'' representations of a random size, or only dimensions equal
to a power of 2. Also the difference between integer and half-integer
(spin) angular momenta is no problem, as we can always ``add'' a
single qubit (a spin 1/2 representation).

\subsubsection{SU(2) or SO(3) representations?}
For integer $\ell$ we have a representation of both SU(2) and SO(3),
while for half-integer angular momentum (thus spin) it is only a
representation of SU(2). Remember that there is a natural 2-to-1
homomorphism from SU(2) to SO(3), thus they are ``almost the
same''. With the spherical harmonics $Y_{\ell m}(\Theta,\phi)$ we have
all SO(3) representations (all odd-dimensional SU(2)
representations). Probably there is a similar representation (in terms
of smooth functions) that directly gives all SU(2)
representations. But spherical harmonics are more familiar to me.

\subsubsection{Symmetric subspace of $N$ qubits.} \label{symm}
There is a nice way to understand what we are trying to do here for
people not familiar with angular momentum (physics) or representations
(of Lie groups). Consider $N$ qubits. Now consider the space of (pure)
states invariant under any permutations of the qubits. This
``symmetric'' subspace is clearly spanned by the ``uniform amplitude
superpositions'' of all binary strings with equal Hamming weight. Thus
one basis state in this symmetric subspace is the superposition (with
equal amplitudes) of all strings with $h$ 1's (and $N-h$ 0's). So for
each Hamming weight we have such a basis state, thus the dimension of
the symmetric subspace is (only) $N+1$.

Now imagine we apply the same transformation to each of the N qubits
(thus overall $U^{\otimes N}$ with $U \in SU(2)$). Note that if we
start with a state in the ``symmetric subspace'' we remain in it, thus
this is an invariant subspace. What we are trying to do in this note
is to ``efficiently'' encode the symmetric subspace in some
$\log_2(N)$ qubits and carry out the transformations corresponding to
doing the same thing to all $N$ original qubits. We already know how
to do transformations corresponding to diagonal SU(2) matrices. To be
able to do everything, it would clearly be sufficient to e.g. do the
transformation corresponding to doing the Hadamard transformation on
each of the $N$ qubits (thus $H^{\otimes N}$). Note that this
``hyper-Hadamard'' can easily be written down by thinking in terms of
the symmetric subspace (use $H^{\otimes N} \ket{x} = 2^{-N/2} \sum_y
(-1)^{x\cdot y} \ket{y}$ and a bit of combinatorics).

What we are doing with the ``symmetric subspace'' of $N$ qubits simply
corresponds to ``adding'' N spin 1/2's in the above sense.

\subsection{Representations of Lie groups other than SU(2).}
It seems that once we know how to ``implement'' SU(2) representations,
we can use this to do any Lie group representations (e.g. SU($n$),
Sp($n$), E$_8$, etc.). This is because in some sense these groups (the
``continuous groups'') can be built up from SU(2).

\subsection{Using this to simulate the chaotic ``kicked top''.}
This investigation was motivated by the idea to simulate the
(quantum-) chaotic ``kicked top'' on a quantum computer (related to me
by Raymond Laflamme and David Poulin, Waterloo). There one considers a
``large spin'' $j$ and repeatedly applies to it a rotation about the
$y$-axis and a diagonal (unitary) transformation of the form $\ket{m}
\to e^{i c \cdot m^2} \ket{m}$. The latter transformation is easy to
implement (as is any rephasing $e^{i f(m)}$ if we can compute $f(m)$,
see e.g. \cite{sim}, eq.(7)). Thus all that is missing is the
$y$-rotation, which is an SU(2) representation.

\subsubsection{Doing it on an NMR quantum computer.}
An interesting simulation of this ``kicked top'' could also be carried
out on an ``ensemble'' quantum computer, thus a quantum computer where
only a constant number of qubits is properly initialised in state
$\ket{0}$ while the rest are in a maximally mixed state. (This model
corresponds to NMR.) The problem is that in our construction, to carry
out computations, we need a lot of properly initialised ancilla
qubits. Thus it would be nice if a future ``nice and natural''
implementation of SU(2) representations could do it without using any
such ancilla qubits (like the Quantum Fast Fourier Transform QFFT), or
then only with a constant number of them.

\subsection{Any connection to the Fourier Transformation?}
The SU(2) representation transformations form a natural class of
unitary transformations, a bit like the Fourier transformations (say
for cyclic groups). In particular we can look at the representation of
the Hadamard transformation (a kind of ``hyper-Hadamard'') and ask
whether it has any connection to the Fourier transformation. We
haven't found any such connection, except maybe in the limit of large
dimensions.

\section{Acknowledgements}
This work was supported in part by the National Security Agency (NSA),
by Advanced Research and Development Activity (ARDA) under the ARDA
Quantum Computing Program, by MITACS and ORDCF.

\end{document}